\documentclass[lettersize]{article}

\usepackage{PRIMEarxiv}

\usepackage[utf8]{inputenc} 
\usepackage[T1]{fontenc}    
\usepackage{hyperref}       
\usepackage{url}            
\usepackage{booktabs}       
\usepackage{amsfonts}       
\usepackage{nicefrac}       
\usepackage{microtype}      
\usepackage{fancyhdr}       
\usepackage{graphicx}       
\graphicspath{{media/}}     
\usepackage{amsmath}
\usepackage{graphicx}
\usepackage{pdflscape}
\usepackage{rotating} 
\usepackage{lmodern}
\usepackage{pdflscape}
\usepackage{lmodern}
\usepackage{array, multirow}

\usepackage{afterpage}
\usepackage{capt-of}
\usepackage{array}
\usepackage{tabularx}
\usepackage{amsmath}
\usepackage{ragged2e}
\usepackage{color}
\usepackage{scalerel}
\newcommand\inclusion{\mathrel{\ThisStyle{\ooalign{$\SavedStyle-$\cr$\SavedStyle<$}}}}
\usepackage{physics}
\usepackage{lscape} 

\title{Machine Learning-Based Assessment of Energy Behavior of RC Shear Walls}

\author{Berkay Topaloglu, Gulsen Taskin Kaya, Ph.D., Fatih Sutcu, Ph.D., Zeynep Tuna Deger, Ph.D.\\
  Istanbul Technical University \\
  \texttt{\{zeynep.tuna@itu.edu.tr (Corresponding author)\}} \\
}

\begin{document}
\maketitle

\begin{abstract}

Current seismic design codes primarily rely on the strength and displacement capacity of structural members and do not account for the influence of the ground motion duration or the hysteretic behavior characteristics. The energy-based approach serves as a supplemental index to response quantities and includes the effect of repeated loads in seismic performance. The design philosophy suggests that the seismic demands are met by the energy dissipation capacity of the structural members. Therefore, the energy dissipation behavior of the structural members should be well understood to achieve an effective energy-based design approach. This study focuses on the energy dissipation capacity of reinforced concrete (RC) shear walls that are widely used in high seismic regions as they provide significant stiffness and strength to resist lateral forces. A machine learning (Gaussian Process Regression (GPR))-based predictive model for energy dissipation capacity of shear walls is developed as a function of wall design parameters (e.g. material properties, reinforcement details). To achieve this, a comprehensive database consisting of 312 shear walls tested under cyclic loading was assembled. Eighteen design parameters are shown to influence energy dissipation, whereas the most important ones are determined by applying sequential backward elimination and by using feature selection methods to reduce the complexity of the predictive model. The ability of the proposed model to make robust and accurate predictions is validated based on novel data with a prediction accuracy (the ratio of predicted/actual values) of around 1.00 and a coefficient of determination (R$^2$) of 0.93. The outcomes of this study are believed to contribute to the energy-based approach by (i) defining the most influential wall properties on the seismic energy dissipation capacity of shear walls and (ii) providing predictive models that can enable comparisons of different wall design configurations to achieve higher energy dissipation capacity.
\end{abstract}

\textit{Keywords: cumulative dissipated energy, energy-based design approach, feature selection, machine learning, predictive models, reinforced concrete shear walls}


\section{Introduction}
\label{S:1}
A new generation earthquake-resistant structural design approach popularly investigated in recent years is the energy-based design approach, which adopts meeting the energy demands of a building by providing sufficient energy capacity by the structural members based on their mechanical and dynamic properties. Housner \cite{housner1958efffct} was one of the first researchers to deal with energy-based structural design. The researcher proposed a correlation of input energy for SDOF systems. McKevitt et al. \cite{mckevitt1980hysteretic} worked on the hysteretic energy spectra in seismic design. Akiyama \cite{akiyama1988earthquake} developed the correlation of input energy. Benavent-Climent \cite{benavent2007energy} investigated an energy-based damage model for steel structures. Whereas some researchers \cite{zahrah1984earthquake,khashaee2003distribution,benavent2010design,okur2012adaptation} examined input energy for SDOF systems, other researchers \cite{goel1997seismic,akbas2001energy,iancovici2008energy,zhang2013effects} investigated input energy for MDOF systems. The design philosophy suggests that enhancing the building design can be achieved either by decreasing the seismic demands or  improving the building’s dynamic characteristics, or both \cite{bertero1992issues}. Therefore, it is essential to understand the energy dissipation behavior of structural members. Experimental evidence and earthquake reconnaissance have shown that shear walls are typically the primary structural members that govern the shear strength and failure of the building, whereas they go undergo plastic deformations (i.e. dissipates energy) in a building \cite{nagae2011design} prior to the columns. In this study, reinforced concrete shear wall properties that influence the energy dissipation capacity are investigated by utilizing machine learning methods. A predictive model is also provided to enable comparing the energy dissipation capacity of shear walls.

Experimental studies conducted in recent years have shown that various shear wall parameters affect energy dissipation capacity as well as shapes of hysteretic curves when subjected to repeated cyclic deformations. For example, structural walls with a lack of transverse reinforcement can lead to pinched hysteretic curves \cite{sengupta2014hysteresis}, whereas energy dissipation capacity can be enhanced by providing higher longitudinal reinforcement ratio and improving the mechanical properties of the rebars \cite{belmouden2007analytical}. Axial load ratio has shown to significantly affect the energy dissipation capacity of shear walls \cite{zhang2000seismic,yun2004behaviour,greifenhagen2005static,su2007seismic,dazio2009quasi,li2015experimental}. Energy dissipation capacity of shear walls is also shown to increase when (i) vertical reinforcement is concentrated at the end of a shear wall \cite{cardenas1973design}, (ii) diagonal reinforcement is used \cite{chiou2004behavior,greifenhagen2005static,shaingchin2007influence,deng2008experimental}, (iii) tie bars are used in the web region \cite{hube2014seismic}, (iv) wall thickness is increased \cite{hube2014seismic}. Liu \cite{liu2004effect} have indicated that concrete compressive strength has an increasing influence on the cumulative dissipated energy which equals to the accumulated energy through the entire test loading protocol, whereas Yan et al. \cite{yan2008seismic} have shown the contrary.  Layysi et al. \cite{layssi2012seismic} observed that Carbon Fiber Reinforced Polymer (CFRP) retrofit increases cumulative dissipated energy.  Besides, other researchers claim that some of the major shear wall parameters do not considerably affect energy capacity. For example, various researchers  have shown that  transverse web reinforcement \cite{sittipunt1995influence,hidalgo2002seismic} and longitudinal web reinforcement \cite{sittipunt1995influence,hidalgo2002seismic} do not significantly influence the energy dissipation capacity. 

An indirect way to estimate the dissipated hysteretic energy under reversed cyclic loading is possible using predictive models. Among analytical models proposed through extensive studies during the past decades \cite{clough1966effect,takeda1970reinforced,saiidi1979simple}, a recent endeavored model has been proposed by Sengupta and Li \cite{sengupta2014hysteresis} based on the Bouc-Wen-Baber-Noori model to predict the hysteretic behavior of reinforced concrete walls, which was calibrated using a database of 100 specimens and derived an analytical model for estimating the hysteretic behavior of reinforced concrete shear walls.

Among data-driven models, machine learning (ML) methods have been used in different structural and earthquake engineering areas in recent years as they have proven effective in providing accurate predictive models. Jeon et al. \cite{jeon2014statistical} studied the shear strength of RC beam-column joints using the conventional multiple linear regression method as well as advanced machine-learning methods of multivariate adaptive regression splines (MARS), and symbolic regression (SR). Zhang et al. \cite{zhang2018machine} utilized machine learning algorithms such as classification and regression tree (CART) and Random Forests to assess structural safety after an earthquake. The same researchers used a dataset consisting of 935 response patterns, 93500 damage patterns, and their corresponding safety states for classification purposes. Davoudi et al. \cite{davoudi2018structural} studied a model that can determine load levels such as shear and moment in slabs and beams by jointly using image processing techniques and machine learning regression methods by utilizing the WEKA toolbox \cite{hall2009weka,Lee2020} in MATLAB for implementing the ML methods and obtained the predictive models of recognizing damage characteristics of the structures. Mangalathu and Jeon \cite{mangalathu2018classification} used classification and regression methods to determine the failure mode of beam-column joints and to predict the shear strength of beam-column joints using a database consisting of 536 experimental tests. Luo and Paal \cite{luo2018machine} proposed a machine learning-based backbone curve model (ML-BCV) for predicting backbone curves of RC columns. Huang and Burton \cite{huang2019classification} classified the failure mode of reinforced concrete infill frames by employing machine learning methods. Siam et al. \cite{siam2019machine} utilized Principal Component Analysis (PCA) as an unsupervised learning algorithm for classifying reinforced masonry shear walls with regard to their failure mode and Projection to Latent Structures (PLS) algorithm as a supervised learning approach for predicting their ultimate drift ratio. More recently, Mangalathu et al. \cite{mangalathu2020data} used machine learning techniques to classify the failure mode of RC shear walls.

The literature review has demonstrated that the impact of major wall design parameters on energy dissipation capacity has been studied, however, no comprehensive investigation has been made to understand the effect of all parameters combined, except Sengupta et al. \cite{sengupta2014hysteresis} who focused on shear walls under monotonic loading. Understanding the energy capacity of shear walls and the effective parameters can lead to more economical and reliable design/assessment, particularly using the energy-based design approach. Based on this motivation, this study aims to predict the cumulative dissipated energy as a function of wall design parameters (e.g. geometric configurations, material properties, and reinforcement details),  referred to as "features" in the machine learning field. To identify the most influential design parameters on the cumulative dissipated energy, a feature selection procedure is performed which reduces the number of design parameters defining the predictive model while reducing its complexity. 

In the following section, a brief background for the normalized cumulative dissipated energy is provided. The reinforced concrete shear wall database assembled to achieve the purposes of this study is introduced in Section \ref{S:3}, whereas an overview of machine learning methods is summarized in Section \ref{S:4}. The input-output configuration of the machine learning problem is discussed in Section \ref{S:5} along with the feature selection methods and associated correlation coefficients. Finally, a GPR-based predictive model verified by the coefficient of determination ($R^2$), mean absolute error (MAE), and root means square error (RMSE) is proposed.

\section{Normalized cumulative dissipated energy of shear walls}
\label{S:2}
Reinforced concrete structural members dissipate energy by undergoing inelastic behavior during cyclic loading. According to many researchers (e.g., Salonikios et al. \cite{salonikios2000cyclic}, Mohamed et al. \cite{mohamed2014experimental}), the energy dissipation capacity of structural members affects the seismic performance of buildings. Therefore, buildings subjected to earthquakes can benefit from shear walls with high energy dissipation capacity within the pre-designated deformation limits. 
The energy dissipation characteristics of RC walls can be examined in various ways, as shown in Eqs. (\ref{eq:1}) and (\ref{eq:2}). The definition of normalized dissipated energy (NDE) according to Hidalgo et al. \cite{hidalgo2002seismic} is given Eq. (\ref{eq:1}),
\begin{align}
\label{eq:1}
\rm{NDE}^{(i)} =\frac{{\rm Energy}\, {\rm dissipated}\, {\rm at}\, {\rm the}\, i^{th}\, {\rm cycle}}{4\bar{V}^{(i)} \bar{\delta }^{(i)} } 
\end{align}
where ${{\bar{V}}^{(i)}}$ is the average of the lateral loads corresponding to maximum load and minimum load at the $i^{th}$ cycle, and ${{\bar{\delta }}^{(i)}}$ is the average of the horizontal displacements corresponding to maximum displacement and minimum displacement at the $i^{th}$ cycle.

The definition of normalized dissipated energy (NDE) according to Kuang and Ho \cite{kuang2008seismic} is provided in Eq.(\ref{eq:2}).

\begin{align}
\label{eq:2}
{\rm NDE}(\mu )=\frac{\int\limits_{0}^{\mu {{\delta }_{y}}}{{V}d\delta }}{{{V}_{i}}{{\delta }_{y}}}
\end{align}
where $\delta $ is lateral top displacement, ${{V}_{i}}$ is the applied load at the top point of the wall when the wall reached its flexural strength,  ${{\delta }_{y}}$ is the average of the horizontal yield displacements corresponding to lateral loads of ${{V}_{i}}$ and $-{{V}_{i}}$, and $\mu$ is the displacement ductility calculated as ${\delta }/{{{\delta }_{y}}}\;$ .

In this study, the energy dissipated by the shear wall at each cycle is calculated as the area enclosed by the load-displacement curve. At the first glance, the cumulative dissipated energy by a specimen throughout the cyclic test might seem an appropriate definition of the energy capacity, however, as different experiments utilize different loading protocols with a different number of cycles at certain steps, this would cause the results be incomparable. To compensate the variety in loading protocols, normalized cumulative dissipated energy (NCDE), calculated as the cumulative dissipated energy divided by the cumulative drift ratio (Eq. (\ref{eq:3})), is considered to be a more appropriate metric to compare specimens subjected to different loading histories. In Eq. (\ref{eq:3}), ${{E}^{(i)}_{{\rm hys}}}$ refers to the dissipated energy at the $i^{th}$ cycle whereas $\Delta^{(i)} $ corresponds to the sum of drift ratios in the positive and negative ends of the $i^{th}$ cycle. It is noted that as elastic deformation range is relatively small, the entire displacement history is used to determine the normalized dissipated energy.
\begin{align}
\label{eq:3}
{\rm NCDE }=\frac{\sum{{{E}^{(i)}_{\rm hys}}}}{\sum{\Delta^{(i)} }}
\end{align}

\section{Experimental database}
\label{S:3}
In the literature, various researchers have established databases for reinforced concrete walls based on their research purposes. A database consisting of 350 RC shear wall specimens tested under cyclic (279) and monotonic loading (71) has been gathered by the University of Patras to develop seismic performance models for RC walls \cite{perus2014series}. Deger and Basdogan \cite{deger2021empiricalshear} created a database of 265 specimens subjected to cyclic loading to propose empirical equations for wall shear strength, whereas Usta et al. \cite{usta2017shear} created a database of 521 shear wall specimens subjected to static (cyclic and monotonic) loading to investigate the sensitivity of shear strength of structural walls under lateral load reversals. More recently, Mangalathu et al. \cite{mangalathu2020data} established a database consisting of 393 specimens to assess the failure mode of shear walls. 

\begin{figure}[!h]
\centering
\includegraphics[scale=0.50]{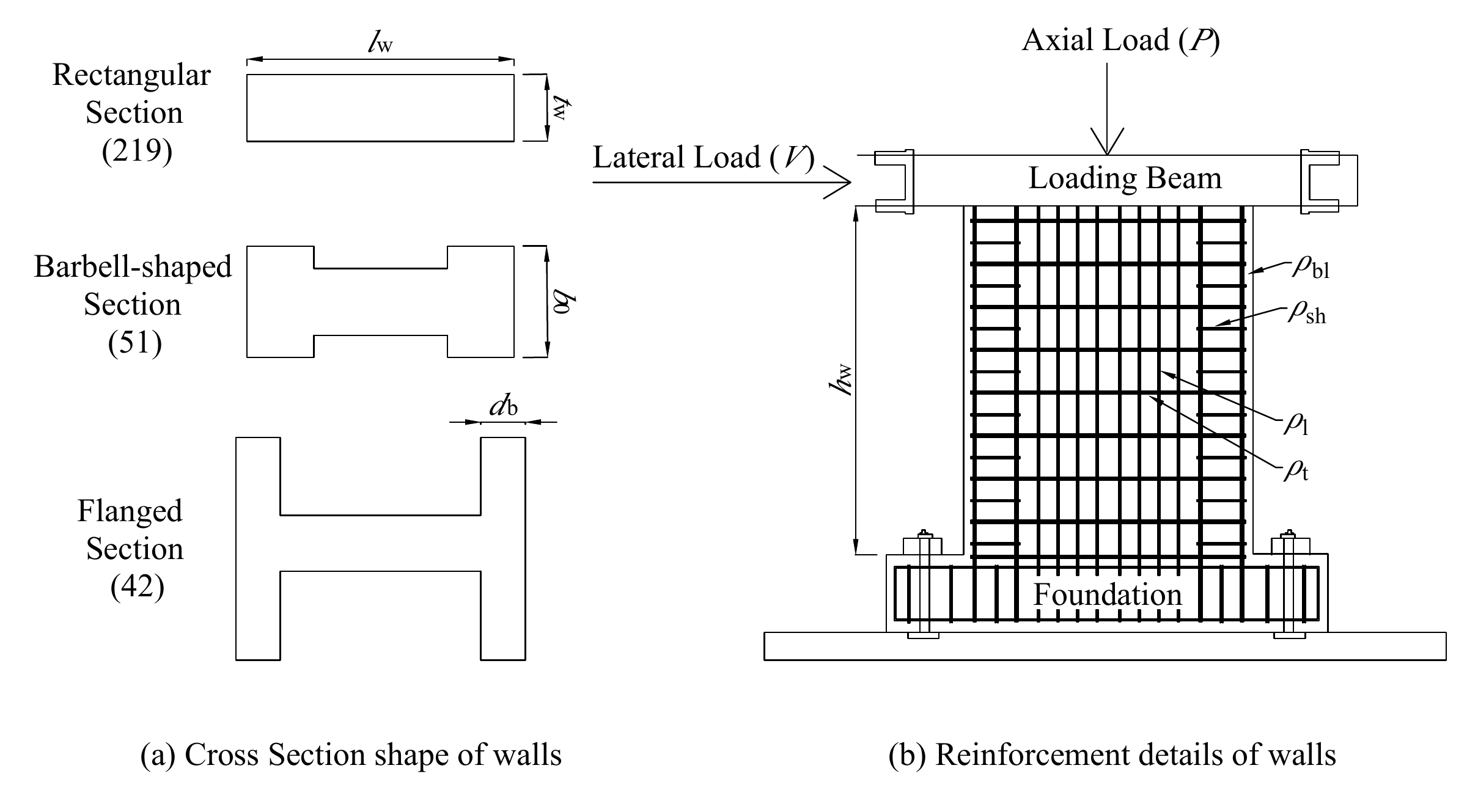}
\caption{RC Shear wall cross section types and geometric properties.}
\label{fig:1}
\end{figure}

Hysteretic load-deformation curves, developed as a result of cyclic loading, are required to calculate the energy dissipation capacity of the shear walls. However, the majority of the existing databases in the literature either include specimens subjected to monotonic loading or lack load-deformation relations of the cyclic tests. To fulfill the needs of this study, a new database is constructed to include 312 specimens subjected to cyclic loading along with their hysteretic curve data. The database includes RC shear walls with rectangular (219/312) and non-rectangular (93/312) cross-sections (Fig. \ref{fig:1}(a)). Shear walls with openings or unsymmetrical specimens are eliminated as energy behavior of such members is relatively unpredictable (i.e., all specimens had symmetrical geometry and symmetrical cyclic loading protocols). Retrofitted specimens and those with diagonal reinforcement are also omitted as such walls are not commonly used in practice, and also because the number of those specimens is not sufficient for the machine learning requirements (i.e., overfitting or underfitting). 

Eighteen design parameters including geometric configurations, material properties, and reinforcement details are used as listed in Table \ref{table:1}. It is noted that for the walls that do not have boundary elements, $b_0 = t_w$ and $d_b = 0$. For the shear walls without stirrups, stirrups spacing is assumed equal to the wall height ($s = h_w$) instead of taking zero ($s = 0$) which would mean infinite number of stirrups. Table \ref{table:1} shows the statistical summary of the eighteen wall design parameters (i.e. features for machine learning tools) for the 312 shear wall specimens. Further details about the database can be obtained elsewhere \cite{tubitak1001report}.

\begin{landscape}
\begin{table}[!h]
\fontsize{1}{1}
\scriptsize
\caption{Design properties of shear walls.}
\centering
\begin{tabular}{ l  l  l  l  l  l  l } 
\hline \hline 
\textbf{No} &  \textbf{Features} &  \textbf{Minimum} &  \textbf{Maximum} &  \textbf{Mean} &  \textbf{Standard deviation} & \textbf{COV (\%)} \\ 
\hline \hline 
 \textbf{1} &  Length (${l}_{w}$) (mm) & 400 & 3500 & 1403 &  602 & 42.89 \\ 
 \textbf{2} &  Height (${h}_{w}$) (mm) & 500 & 12000 & 2173 & 1199 & 55.18 \\ 
 \textbf{3} &  Thickness (${t}_{w}$) (mm) & 26 & 300 & 131 & 45 & 34.39 \\ 
 \textbf{4} &  Concrete compressive strength ($f_{c}^{'}$)(MPa) & 12.35 & 117.00 & 40.59 & 19.67 & 48.46 \\ 
 \textbf{5} & Transverse web reinforcement yield strength (${f}_{yt}$)(MPa) & 216.00 & 1001.00 & 491.45 &  131.07 & 26.67 \\  
 \textbf{6} &  Transverse boundary reinforcement yield strength (f${}_{ysh}$)(MPa) & - & 1262.00 & 324.38 &  273.18 & 84.22 \\ 
 \textbf{7} &  Vertical web reinforcement yield strength (${f}_{yl}$)(MPa) & 216.00 & 1001.00 & 499.26 &  129.33 & 25.90 \\ 
 \textbf{8} &  Vertical boundary reinforcement yield strength (${{f}_{ybl}}$)(MPa) & - & 1450.80 & 451.86 &  186.26 & 41.22 \\ 
 \textbf{9} &  Transverse web reinforcement ratio (${{\rho }_{t}}$)(\%) & 0.11 & 2.42 & 0.56 &  0.32 & 57.14 \\ 
 \textbf{10} &  Transverse boundary reinforcement ratio (${{\rho }_{sh}}$)(\%) & 0.00 & 9.57 & 0.68 &  0.88 & 129.41 \\ 
 \textbf{11} &  Vertical web reinforcement ratio (${{\rho }_{l}}$)(\%) & 0.13 & 3.29 & 0.57 &  0.38 & 66.67 \\ 
 \textbf{12} &  Vertical boundary reinforcement ratio (${{\rho }_{bl}}$)(\%) & 0.00 & 13.04 & 3.33 &  3.05 & 91.59 \\ 
 \textbf{13} &  Axial Load Ratio (${P}/{({{A}_{g}}} f_{c}^{'})$) & 0.00 & 0.50 & 0.10 &  0.12 & 114.45 \\ 
 \textbf{14} &  Boundary element depth (${b}_{0}$)(mm) & 50.00 & 1500.00 & 206.57 &  160.97 & 77.93 \\  
 \textbf{15} &  Boundary element length (${d}_{b}$)(mm) & 0.00 & 590.80 & 176.07 &  99.45 & 56.48 \\ 
 \textbf{16} &  Hoop spacing / Boundary element length (${s}/{d}_{b}$) & 0.00 & 52.08 & 3.93 &  7.38 & 187.79 \\  
 \textbf{17} &  Aspect ratio ($AR$) & 0.33 & 7.38 & 1.67 &  0.79 & 47.31 \\ 
 \textbf{18} &  Shear span ratio (${M}/{Vl}_{w}$) & 0.33 & 7.38 & 1.75 &  0.91 & 52.00 \\ \hline
\end{tabular}
\label{table:1}
\end{table}
\end{landscape}

\subsection{Estimation of the dissipated hysteresis energy using load-displacement diagrams}
\begin{figure}[!h]
\centering
\includegraphics[width=0.85\columnwidth]{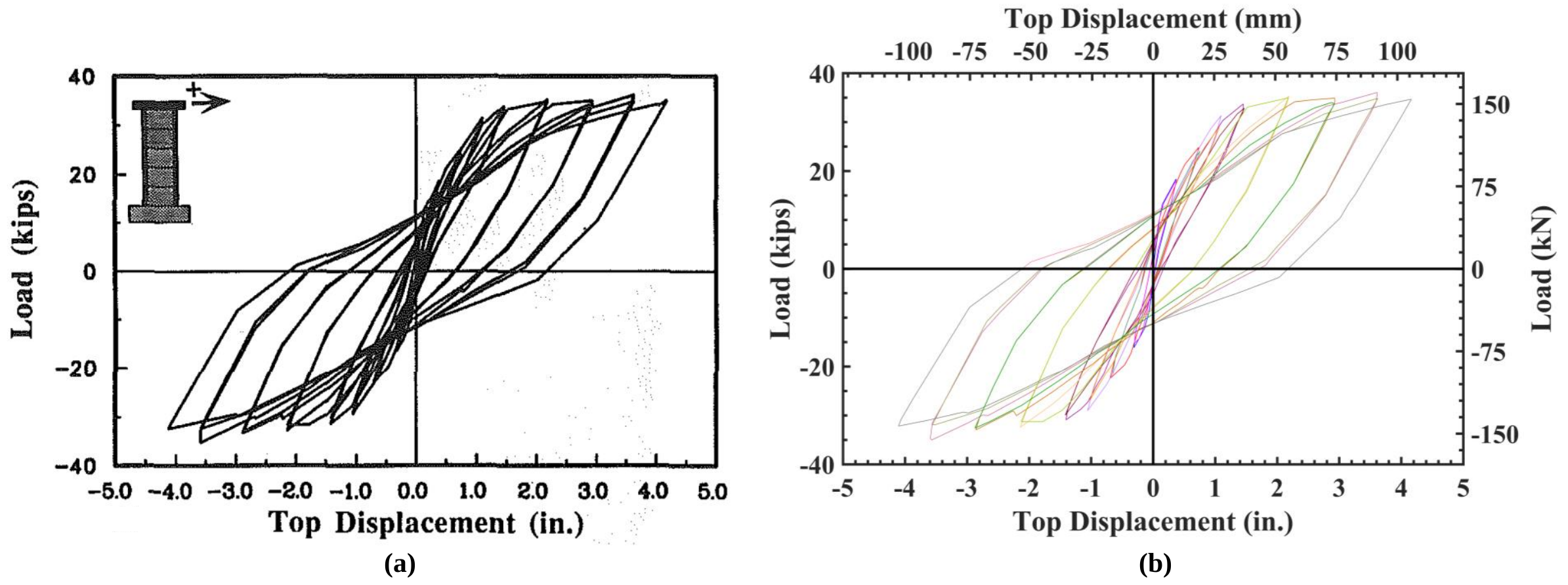}
\caption{Load-deformation curve for specimen W1 \cite{ali1991rc}: (a) Original plot (b) Digitized plot.}
\label{fig:2}
\end{figure}

A manual digitization procedure is developed to digitally create hysteretic curves for shear wall specimens that were provided with images, and digital data of which was unavailable. The digitization procedure includes hand-tracing in AutoCAD and transferring the data to MATLAB environment. Fig. \ref{fig:2} shows an example of the digitized hysteretic curve for a shear wall specimen. To verify the accuracy of the procedure, manually digitized hysteretic curve data were compared to the digital test data for 40 arbitrary specimens with available digital data, verifying that the human error was negligible ($<5\%$). 

\begin{figure}[!h]
\centering
\includegraphics[width=0.50\columnwidth]{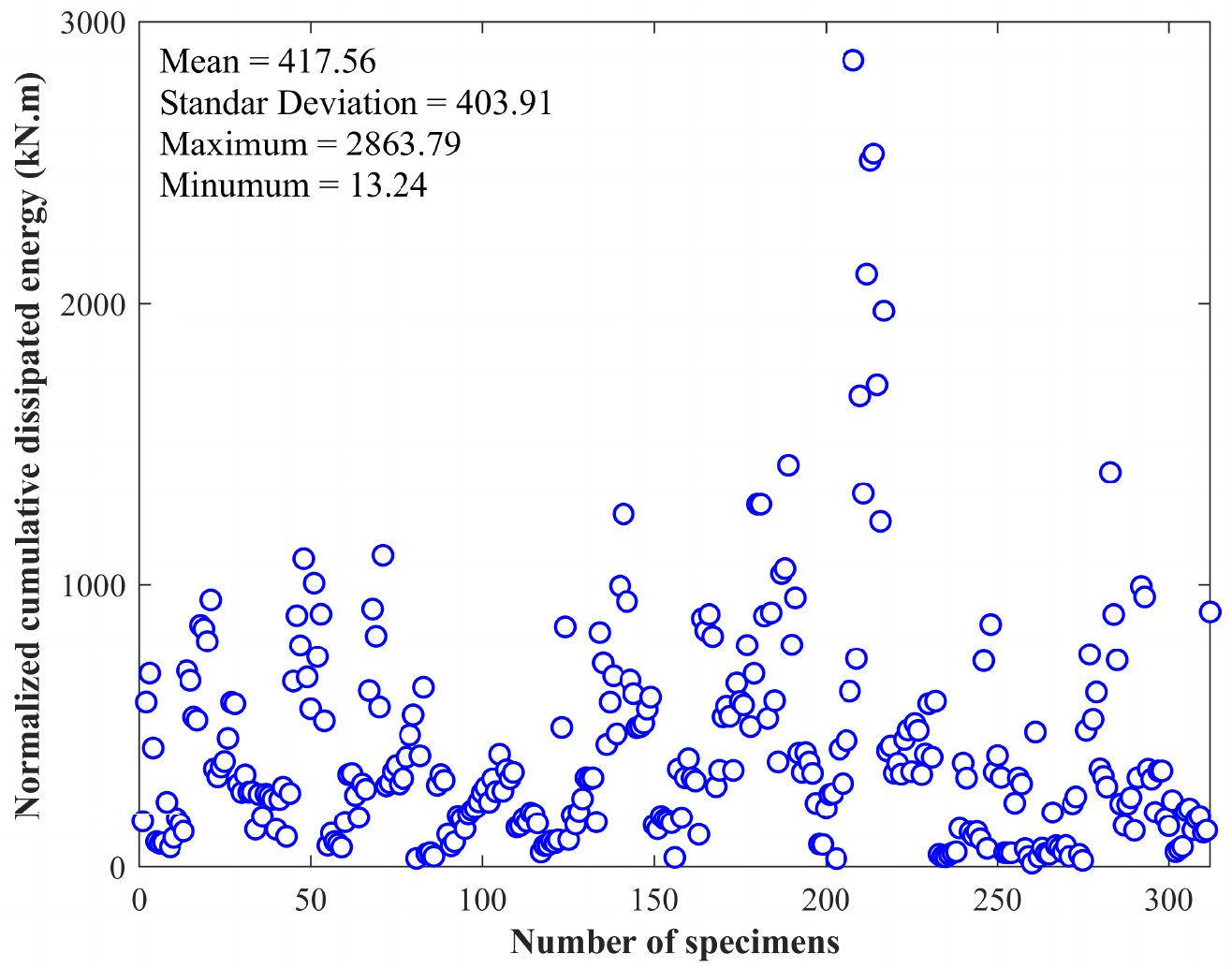}
\caption{Normalized cumulative dissipated energy values.}
\label{fig:3}
\end{figure}

A total of 131 specimens are manually digitized, whereas digital test data are obtained from the University of Patras FP7 Program \cite{perus2014series} data portal (107 specimens) and from relevant researchers (74 specimens). The normalized cumulative dissipated energy is calculated for each of the 312 shear walls based on Eq. (\ref{eq:3}), whereas their distribution is presented in Fig. \ref{fig:3} along with a summary of their statistics.

\section{Overview of machine learning  methods}
\label{S:4}
Feature selection is an important machine learning method aiming to rank the features based on specific predefined criteria regarding their contribution to the prediction function \cite{lee2010computerized,shang2018imputation}.  This study aims to select the best subset out of eighteen features (i.e. design parameters) that strongly influence estimating the NCDE of the shear walls. Four feature selection methods were used, namely: Linear Regression (LR), Lasso Regression (LASSO), Neighborhood Component Analysis (NCA), and Gaussian Process Regression (GPR), each of which is briefly summarized in the following subsections.  Note that some of these algorithms are directly designed as a feature selection method, whereas some are regression methods in which the coefficients of the prediction function are evaluated as a criterion to rank the features with respect to their contribution to the energy of the shear walls. As each feature in the database has a different unit (e.g. mm, kN), all features (for both training set and test set) are scaled to the range [-1,1] using min-max values of training set so that the performance of the machine learning method is less affected. Feature subsets are generated based on the recommendations made by the corresponding feature selection method, whereas the subset achieving the highest performance in estimating the NCDE is selected. All analyses are implemented in MATLAB by utilizing  Statistics and Machine Learning Toolbox™ \cite{Mathwork2020}. 

\subsection{Linear regression}
Let $ X=\left\{ x_1, \cdots, x_n \right\} \inclusion {{\mathbb{R}}^{m}}$ be a set of $m$ number of specimens and $n$ number of design parameters in the associated database and $f(\mathrm{x}):{{\mathbb{R}}^{n}}\to \mathbb{R}$ be a prediction function. Note that the NCDE of the shear walls is denoted as ${{y}_{i}}$ for each $i^{th}$ specimen in the database.

Linear regression is the simplest technique providing a linear model, which is constituted based on the linear relationship between input and output of a system \cite{friedman2001elements}. The prediction function has the following form: 
\begin{equation}
\label{eq:4}
f({{x}_{1}},\,\cdots ,\,{{x}_{n}})={{\beta }_{0}}+{{\beta }_{1}}{{x}_{1}}+\cdots +{{\beta }_{n}}{{x}_{n}}
\end{equation}
where ${{\beta }_{i}}$ refers to  the coefficients of the $i^{th}$ feature and indicates the contribution of the $i^{th}$ feature on the model output.  The empirical risk function in linear regression is defined as follows: 
\begin{equation}
\label{eq:5}
\mathcal{L}(\beta)= \sum_{i=1}^m \left(f(x)_i-y_i\right)^2
\end{equation}

The coefficients are determined as in the following equation once the above  function is minimized, leading to solving the following systems of linear equations.  
\begin{equation}
\label{eq:6}
\hat{\beta }={{({{\mathbf{X}}^{T}}\mathbf{X})}^{-1}}{{\mathbf{X}}^{T}}\mathbf{y}
\end{equation}
where $\mathbf{X}$ is the-so-called $m\times n$ design matrix, and $\mathbf{y}$ is a $m$-dimensional vector holding all the outputs of NCDE. Each coefficient can be regarded as a criterion for feature selection, meaning that the larger the value, the more important the feature is. 

\subsection{Least absolute shrinkage and selection operator (LASSO) regression}
Lasso regression is linear regression formula that includes an LASSO penalty. Lasso penalty has an effect as a regularizer, shrinking the coefficients and providing a sparse solution. The Lasso estimate ${{\hat{\beta }}_{lasso}}$, is defined as follows:

\begin{equation}
\label{eq:7}
{{\hat{\beta }}_{lasso}}=\underset{\beta }{\mathop{\arg \min }}\,\left[ \sum_{i=1}^m \left(f(x)_i-y_i\right)^2+\lambda \sum\limits_{j=1}^{n}{ \abs{{\beta }_{j}} } \right]
\end{equation}
The second term in Eq. (\ref{eq:7}) plays an a vital role in providing a sparse solution as well as preventing overfitting, meaning that the irrelevant features having zero-coefficients are removed from the input space, hence the LASSO can be utilized as a feature selection method. The most significant features  are retained while the insignificant features  are removed from the model.

\subsection{Neighborhood component analysis}
NCA is specially designed to improve the performance of the KNN and therefore uses a stochastic neighborhood assignment by maximizing the performance of the leave-one-out (LOO) regression error on the training set   \cite{goldberger2004neighbourhood}. For each sample, ${{\mathbf{x}}_{i}}$, the following probability distribution is defined to choose the reference nearest neighborhood.  

Eq. (\ref{eq:8}) gives margin that is expressed in terms of weighing vector $\mathbf{w}$, whereas the margin corresponds to the distance between two samples ${{\mathbf{x}}_{i}}$ and ${{\mathbf{x}}_{j}}$  \cite{yang2012neighborhood}.
\begin{equation}
\label{eq:8}
D\mathbf{w}\left( {{\mathbf{x}}_{i}},{{\mathbf{x}}_{j}} \right)=\sum\limits_{l=1}^{n}{w_{1}^{2} \abs{{{x}_{il}}-{{x}_{jl}}} }
\end{equation}
where ${{w}_{l}}$ is the $l^{th}$ feature weight. For a given training set $S=\left\{ (\mathbf{x_i}, y_i), i=1,\cdots,m \right\}$, the method starts with randomly selecting a reference point ($\mathbf{x_j}$) from $S$ for the sample $\mathbf{x}$. If the LOO is applied, the randomized regression model estimates the output for $\mathbf{x_i}$ using other data excluding the reference sample $(\mathbf{x_i},y_i)$ in the training set. Given a data point ${{\mathbf{x}}_{i}}$, the probability of drawing ${{\mathbf{x}}_{j}}$ as a reference point of $\mathbf{x_i}$ is obtained as following:

\begin{align}
\label{eq:9}
{{p}_{ij}}=\left\{ \begin{array}{*{35}{l}}
   \frac{k\left( D\mathbf{w}\left( {{\mathbf{x}}_{i}},{{\mathbf{x}}_{j}} \right) \right)}{\sum\nolimits_{k\ne i}{k\left( D\mathbf{w}\left( {{\mathbf{x}}_{i}},{{\mathbf{x}}_{j}} \right) \right)}}, & i\ne j  \\
   0, & i=j  \\
\end{array} \right.
\end{align}
In Eq. (\ref{eq:9}), the $k\left( D\mathbf{w}\left( {{x}_{i}},{{x}_{j}} \right) \right)$ is a kernel function. The ${{\hat{y}}_{i}}$ is the prediction value of the randomized regression model, whereas ${{y}_{i}}$ is the actual response value for ${{x}_{i}}$. The loss function measuring the disagreement between ${{\hat{y}}_{i}}$ and ${{y}_{i}}$, is defined as $\mathcal{L}\left( {{y}_{i}},{{{\hat{y}}}_{i}} \right)=\abs{ {{y}_{i}}-{{{\hat{y}}}_{i}}}$. The average value of $\mathcal{L}\left( {{y}_{i}},{{{\hat{y}}}_{i}} \right)$ is
\begin{eqnarray}
\label{eq:10}
{{\mathcal{L}}_{i}} &=&E\left( \mathcal{L}\left( {{y}_{i}},{{{\hat{y}}}_{i}} \right)\mid {{S}^{-i}} \right) \nonumber\\
&=&\sum\limits_{j=1,\,j\ne i}^{m}{{{p}_{ij}}\mathcal{L}({{y}_{i}},{{y}_{j}})}
\end{eqnarray}
The objective function $f(\mathbf{w})$ is minimized by adding regularization term $\lambda$ in Eq. (\ref{eq:11}).
\begin{equation}
\label{eq:11}
f(\mathbf{w})=\frac{1}{m}\sum\limits_{i=1}^{m}{{{\mathcal{L}}_{i}}+\lambda \sum\limits_{r=1}^{n}{w_{r}^{2}}}
\end{equation}

\subsection{Gaussian process regression}
 Gaussian process regression (GPR) is a multivariate nonparametric kernel method, which can be utilized both classification and regression purposes \cite{williams2006gaussian,kang2020displacement}. 

The linear correlation between the $\mathbf{x}$ and the $y$ can be defined as
\begin{equation}
\label{eq:12}
{y_i}=f({{\mathbf{x}}})_i+\varepsilon 
\end{equation}
where $f(\mathbf{x})$ is the basis regression function, and $\varepsilon $ is the residue term drawn from $N(0,\,{{\sigma }^{2}})$. A GPR model represents the relation between input and output by including latent variables derived from a Gaussian process (GP), and basis functions, $f$. A GP includes random variables that have a joint Gaussian distribution. The  prediction is found by a covariance function of the latent variables, and  the inputs $x$ is projected into a $p$-dimensional feature space by the basis function. The $h(\mathbf{x})$ is assessed as a random variable. A GP consists of mean and covariance functions that are defined as follows:

\begin{eqnarray}
   E(h(\mathbf{x})) &=&m(\mathbf{x}) \\
   Cov\left[ h(\mathbf{x}),h(\mathbf{{x}'}) \right]&=&E\left[ \left\{ h(\mathbf{x})-m(\mathbf{x}) \right\}\left\{ h(\mathbf{{x}'})-m(\mathbf{{x}'}) \right\} \right]=k(\mathbf{x},\mathbf{{x}'})
\end{eqnarray}

Considering the model in Eq. (\ref{eq:13}), $h(\mathbf{x})\sim GP(0,k(\mathbf{x},\mathbf{{x}'}))$, and  $h(\mathbf{x})$ corresponds to from a zero mean GP with covariance function. The $f(\mathbf{x})$ transforms the original feature vector $\mathbf{x} \in \mathbb{R}^{n}$  into a new feature vector $f(\mathbf{x}) \in \mathbb{R}^{p}$. 
\begin{equation}
\label{eq:13}
y=f(\mathbf{x}) + h(\mathbf{x})
\end{equation}
The predictive function $\mathbf{y}$ can be defined as
\begin{equation}
\label{eq:14}
\mathbf{P}\left( {{\mathbf{y}}_{i}}\mid h({{\mathbf{x}}_{i}}),{{\mathbf{x}}_{i}} \right)\sim \mathbf{N}\left( {{\mathbf{y}}_{i}}\mid f{{({{\mathbf{x}}_{i}})}} +h({{\mathbf{x}}_{i}}),{{\sigma }^{2}} \right)
\end{equation}
As the GPR is a probabilistic model, latent variables $h({{\mathbf{x}}_{i}})$ corresponding each features $x_i$ makes the GPR model a nonparametric. A vectorized form is defined as follows:

\begin{equation}
\label{eq:15}
\mathbf{P}\left( \mathbf{y}\mid \mathbf{H},\mathbf{X} \right)\sim \mathbf{N}\left( \mathbf{y}\mid \mathbf{f} +\mathbf{H},{{\sigma }^{2}}\mathbf{I} \right)
\end{equation}
Eq. (\ref{eq:16}) shows the joint distribution of latent variables, $h(\mathbf{x_1}),\cdots,h(\mathbf{x_m})$.
\begin{equation}
\label{eq:16}
\mathbf{P}\left( \mathbf{H}\mid \mathbf{X} \right)\sim \mathbf{N}\left( \mathbf{H}\mid 0,\mathbf{K}(\mathbf{X},\mathbf{X}) \right)
\end{equation}
$\mathbf{K}(\mathbf{X},\mathbf{X})$ in Eq. (\ref{eq:16}), is defined as:
\begin{equation}
\label{eq:17}
\mathbf{K}(\mathbf{X},\mathbf{X})=\left[ \begin{matrix}
   k({{\mathbf{x}}_{1}},{{\mathbf{x}}_{1}}) &  \cdots  & k({{\mathbf{x}}_{1}},{{\mathbf{x}}_{m}})  \\
   
   \vdots   & \vdots  & \vdots   \\
   k({{\mathbf{x}}_{m}},{{\mathbf{x}}_{1}}) &  \cdots  & k({{\mathbf{x}}_{m}},{{\mathbf{x}}_{m}})  \\
\end{matrix} \right]
\end{equation}
To define the covariance function $k(\mathbf{x},\mathbf{{x}'})$, various kernel functions can be used, and  the covariance function is generally parameterized in terms of the kernel parameters in vector $\theta$:


\begin{equation}
\label{eq:20}
k\left( {{\mathbf x}_{i}},{{\mathbf x}_{j}}\mid \theta  \right)=\sigma _{f}^{2}\exp \left[ -\frac{1}{2}\sum\limits_{m=1}^{m}{\frac{{{\left( {{\mathbf x}_{im}}-{{\mathbf x}_{jm}} \right)}^{2}}}{\sigma _{m}^{2}}} \right]
\end{equation}
where $\sigma_m$ represents the length scale for features $m$, and $\sigma_f$ is the signal standard deviation. The hyperparameters $\theta$ is
\begin{equation}
\begin{matrix}
  {{\theta }_{m}}=\log \sigma_m , & \text{for }m=1,\cdots ,m  \\
  {{\theta }_{d+1}}=\log \sigma_f   \\
\end{matrix}
\end{equation}
Initial kernel parameter values and initial noise standard deviation is determined to create a GPR model, then the model is created with an automatic relevance determination (ARD) squared exponential kernel function in MATLAB. To find the feature weights ($w$), the exponential of the negative learned length scales are taken. The feature weights are normalized as shown in Eq.(\ref{eq:normalizedw}) and are used for the feature selection process.
\begin{equation}
\label{eq:normalizedw}
{{w}_{normalized}}=\frac{w}{\sum\limits_{i=1}^{n}{{{w}_{i}}}}
\end{equation}


\section{Experimental analysis and results}
\label{S:5}
The database is randomly split into the training (80$\%$) and test sets (20$\%$) to investigate the effect of the design parameters on the NCDE and find the most accurate predictive model with the least number of features possible. The training data set is utilized to develop the model that predicts the output using the aforementioned machine learning methods, whereas the test data set prevents overfitting thanks to being unseen (new) data that the model encounters for the first time, thus helping to find the most reliable predictive model. A thousand different training-test set combinations are randomly created to ensure reliable and robust results. Five statistical metrics are used to assess the accuracy of the model based on the test dataset, namely: mean absolute error (MAE), root mean square error (RMSE), relative RMSE (RELRMSE), coefficient of determination (R$^2$), and mean prediction accuracy (the ratio of predicted to actual NCDE values) as given in Eqs. (\ref{eq:22}) - (\ref{eq:26}).
\begin{equation}
\label{eq:22}
MAE=\frac{1}{m}\sum\limits_{i=1}^{m}{ \abs{{{{\hat{y}}}_{i}}-{{y}_{i}}} }
\end{equation}

\begin{equation}
\label{eq:23}
RMSE=\sqrt{\frac{1}{m}\sum\limits_{i=1}^{m}{{{\left( {{{\hat{y}}}_{i}}-{{y}_{i}} \right)}^{2}}}}
\end{equation}

\begin{equation}
\label{eq:24}
RELRMSE=\frac{RMSE}{\left( \frac{1}{m}\sum\limits_{i=1}^{m}{{{{\hat{y}}}_{i}}} \right)}
\end{equation}

\begin{equation}
\label{eq:25}
{{R}^{2}}={{\left( \frac{\operatorname{cov}(\mathbf{y},\mathbf{\hat{y}})}{{{\sigma }_{\mathbf{y}}}{{\sigma }_{{\mathbf{\hat{y}}}}}} \right)}^{2}}
\end{equation}

\begin{equation}
\label{eq:26}
PA=\frac{1}{m}\sum\limits_{i=1}^{m}{\frac{{{{\hat{y}}}_{i}}}{{{y}_{i}}}}
\end{equation}

\subsection{Data transformation}

\begin{figure}[!h]
\centering
\includegraphics[width=0.55\columnwidth]{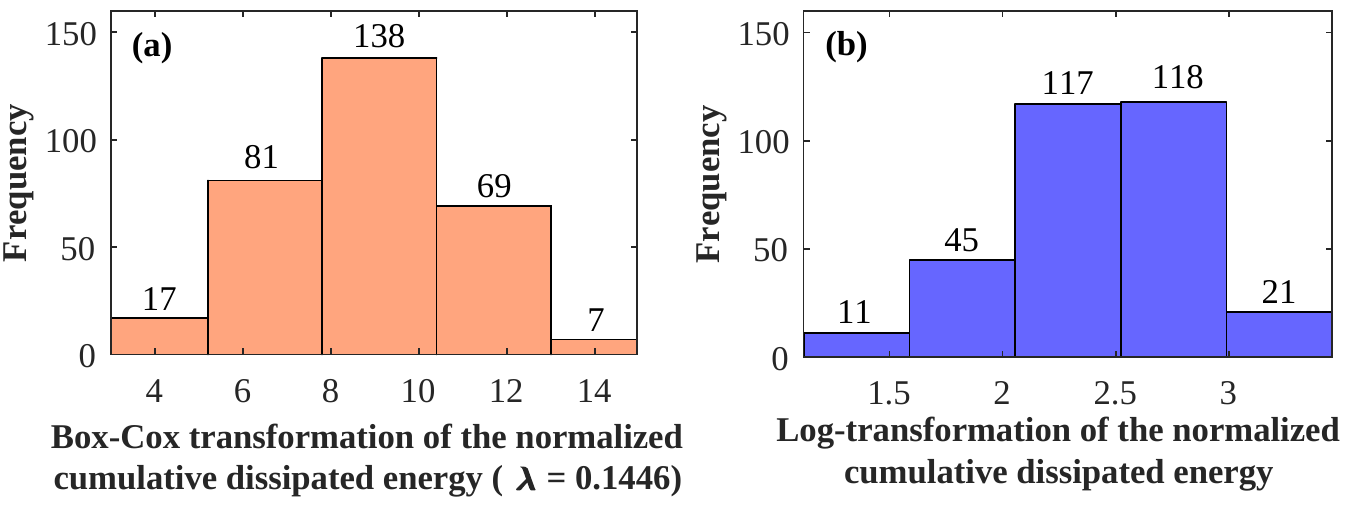}
\caption{Distribution of NCDE values for (a) Box-Cox transformation (optimal $\lambda = 0.1446$), (b) Log transformation.}
\label{fig:4}
\end{figure}

The accuracy of the predictive model provided by the machine learning methods can be improved by simply transforming the data into a more informative format. In this context, data pre-processing techniques have proven to be a proper technique for improving the quality of the results \cite{jiang2008can}. Among several data transformation techniques, since Box-cox transformation and log transformation were efficient, such transformations are applied to the outputs. The original form of the Box-Cox transformation equation is presented in Eq. (\ref{eq:27}) \cite{box1964analysis}.

\begin{equation}
\label{eq:27}
y(\lambda )=\left\{\begin{array}{c} {\begin{array}{cc} {\frac{y^{\lambda } -1}{\lambda } ,} & {{\rm if}\lambda \ne 0} \end{array}} \\ {\begin{array}{cc} {\log (y),} & {{\rm if}\lambda =0} \end{array}} \end{array}\right. 
\end{equation}
where $y$ is the data value associated with the output, and $\lambda$ is a value between -5 and 5. The optimal $\lambda$ is the one that transforms the dataset to the best approximation of normal distribution.  Fig. \ref{fig:4} demonstrates the Box-Cox (optimal $\lambda = 0.1446$) and log transformations of the NCDE. The use of a log transformation or Box-Cox transformation on the output is observed to increase the accuracy of the predictive model of each method. Table \ref{table:2} presents the effect of data transformation techniques on the GPR model results.

\begin{table}[h!]
\fontsize{6}{6}
\caption{Effect of data transformation techniques on GPR model results (1000 repeats).}
\centering
\begin{tabular}{l l l l l} 
 \hline \hline 
 \textbf{METHOD} & \textbf{Data Transformation} &  \textbf{The Best R${}^{2}$ } &  \textbf{Mean R${}^{2}$ } & \textbf{Min. \newline  RELRMSE} \\ [0.5ex] 
 \hline\hline
  GPR &  No transformation &  0.870 &  0.535 & 0.38 \\ 
  GPR &  Log transformation &  \textbf{0.954} &  0.811 & \textbf{0.22} \\
  GPR &  Box-Cox transformation &  0.954 &  \textbf{0.815} & 0.23 \\ [0.5ex] 
 \hline
\end{tabular}
\label{table:2}
\end{table}

\begin{figure}[h]
\centering
\includegraphics[width=0.9\columnwidth]{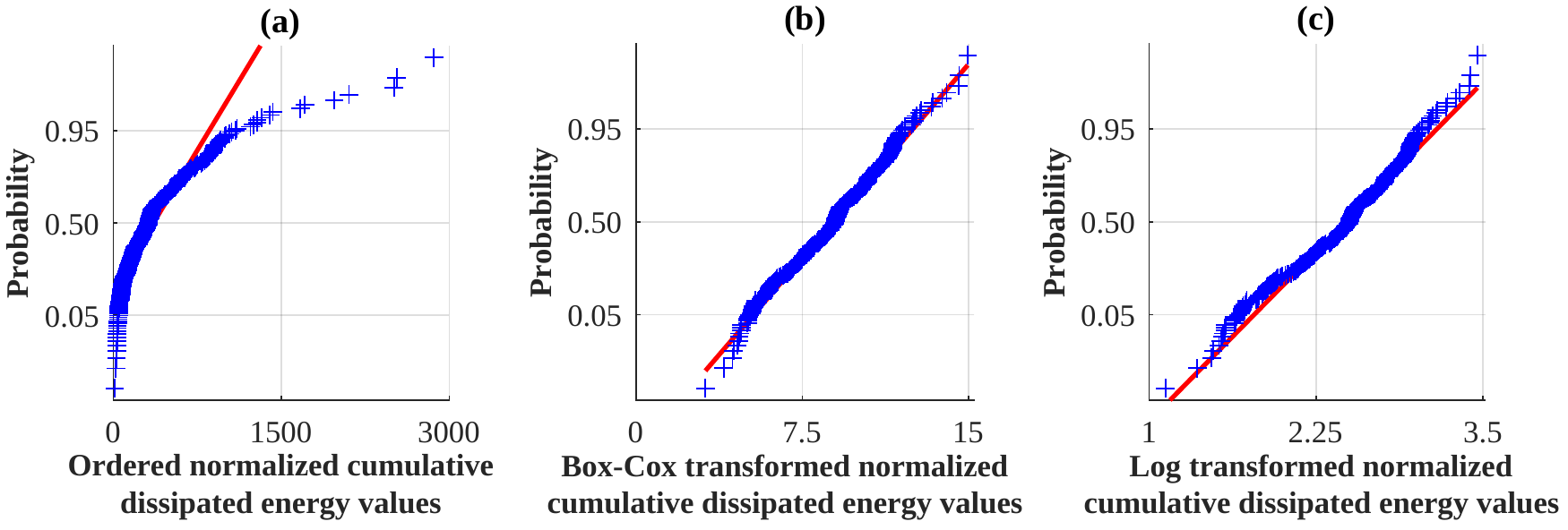}
\caption{Comparing actual and transformed normalized cumulative dissipated energy values  in terms of Gaussian probability plots.}
\label{fig:5}
\end{figure}

The Box-Cox transformation converts the distribution of output values to better fit the normal distribution, which can be clearly observed in Fig. \ref{fig:5} where normal probability plots of the output before and after Box-Cox transformation are presented with uniform order statistics medians approximated as suggested by Filliben \cite{filliben1975probability} (Eq. (\ref{eq:28})).

\begin{equation}
\label{eq:28}
m_{i} =\left\{\begin{array}{ll} {1-m_{m} } & {i=1} \\ {{(i-0.3175)\mathord{\left/ {\vphantom {(i-0.3175) (m+0.365)}} \right. \kern-\nulldelimiterspace} (m+0.365)} } & {i=2,\cdots ,\, m-1} \\ {0.5^{\frac{1}{m} } } & {i=m} \end{array}\right. 
\end{equation}
In Eq. (\ref{eq:28}), ${{m}_{i}}$ values are normal order statistic medians; $m$ is number of specimens, and ${{m}_{m}}$ value refers to the normal order statistic median of the $m^{th}$ feature. The normal distribution whose graph resembles a bell curve can be explained by two basic parameters: the mean value and the variance. The normal distribution is symmetric around the mean value, and the data near the mean value reveals more frequently than the data far from the mean depending on the small variance value. Fig. \ref{fig:5} shows that the red line represents the normal distribution while the blue points refer to the NCDE values. If the data follow a normal distribution, then the blue points are expected to close the red line. Likewise, the distribution other than the Gaussian introduces curvature in the data plot. It is observed that the transformed data points in Fig. \ref{fig:5}(b) and Fig. \ref{fig:5}(c) show less departure from the straight line, representing a closer distribution to the theoretical normal distribution compared to the original data (Fig. \ref{fig:5}(a)). Though Box-cox transformation and log transformation showed similar distribution, log transformation is preferred due to its practical use. 

\subsection{Feature selection analysis}

\begin{figure}[!h]
\centering
\includegraphics[width=0.5\columnwidth]{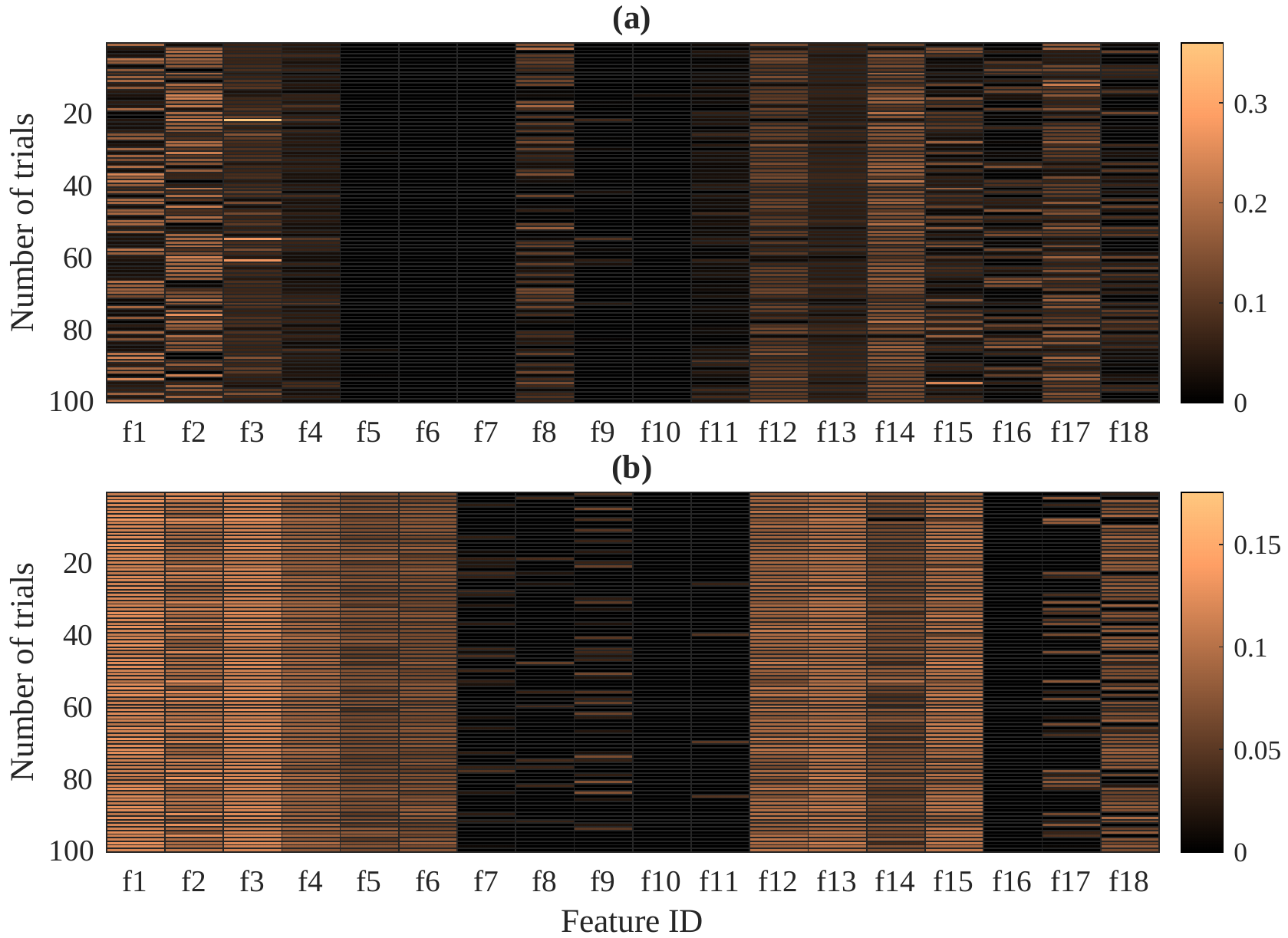}
\caption{Feature weights calculated by (a) GPR and (b) NCA based on 100 random trials with different training and testing data sets.}
\label{fig:6}
\end{figure}

\begin{figure}[!h]
\centering
\includegraphics[width=0.8\columnwidth]{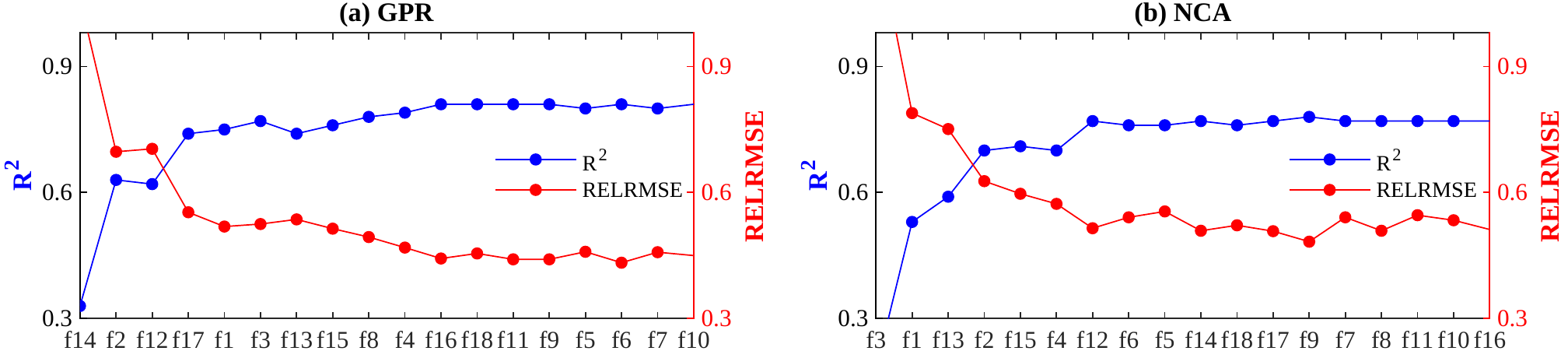}
\caption{Mean values of R$^2$ and RELRMSE results of the analysis of (a) GPR, and (b) NCA methods after feature selection processing for 100 trials.}
\label{fig:heatmapsR2}
\end{figure}

An important aspect of this paper is to obtain the most influential shear wall design parameters by (i) ranking them and (ii) applying sequential backward elimination (SBE) with feature selection methods. To rank the importance of the features, feature weights are computed using the feature selection methods mentioned in Section \ref{S:4}.  Because each method reveals different feature subsets, the results are evaluated individually.  A hundred trials are performed for each method to enhance reliability and robustness in the feature rankings.  The mean R$^ 2 $ values of the hundred trials are 0.445, 0.405, 0.768, and 0.811 for LR, LASSO, NCA, and GPR, respectively.  Linear methods are not assessed in terms of feature weights as they are unsuccessful to create an accurate predictive model. Fig. \ref{fig:6} shows the weights of eighteen features as a heat map for each trial for the GPR and NCA methods. The x-axis indicates the feature ID numbers in consistence with Table \ref{table:1}, whereas y-axes presents results of each trial. Note that the heat map colors become lighter as value of the feature weight increases. That is, the lightest colored feature is ranked the first, whereas the darkest colored feature is ranked the last. To understand the effect of including each feature in the prediction of the NCDE, the ranked features are introduced to the nonlinear methods one at a time in the order of their descending ranking value (starting with the highest feature weight). Fig. \ref{fig:heatmapsR2} shows how the number of ranked features at each run of NDCE estimation affects the prediction results in terms of $ R ^ 2 $  and RELRMSE. The results indicate that the GPR and NCA methods are able to capture the best performance and become stable when the number of features reach eleven and seven, respectively.

\begin{figure}[!h]
\centering
\includegraphics[width=0.8\columnwidth]{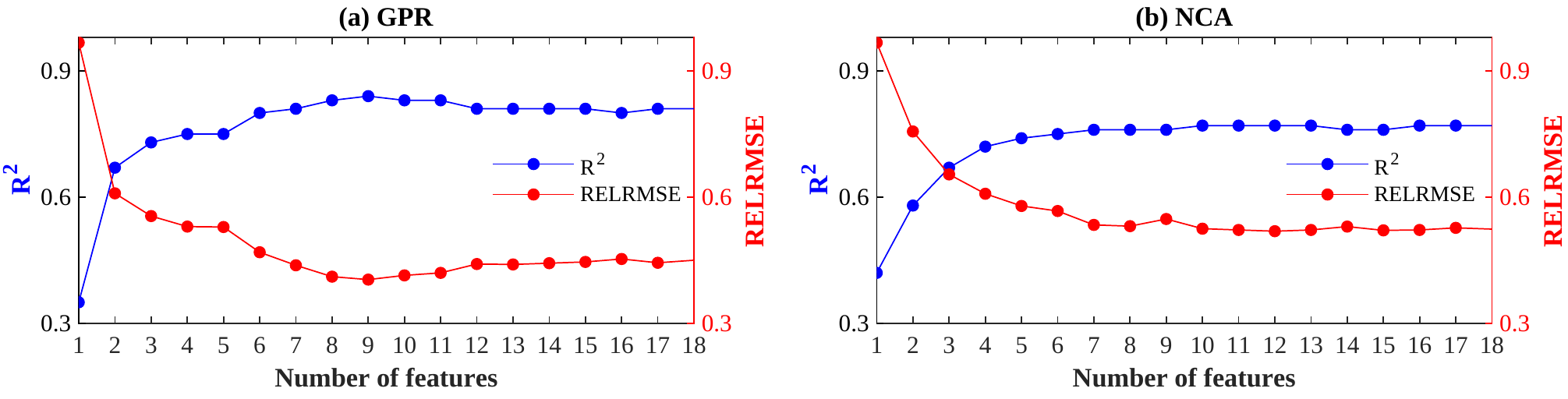}
\caption{Mean values of R$^2$ and RELRMSE results of the analysis of the SBE according to (a) GPR and (b) NCA for 1000 trials.}
\label{fig:8}
\end{figure}

Besides the regression-based feature selection method, sequential backward elimination (SBE) is used to find the best subset that includes as few features as possible. To perform feature selection with SBE, starting from the full candidate set, the features are removed until the most relevant subset is obtained while using nonlinear regression methods. The subsets are created by reducing the number of features of the candidate set one by one at each step, and each subset is examined in terms of  R$^2$ and RELRMSE. This process is repeated until the best performance is achieved, the results of which are presented in Fig. \ref{fig:8} in terms of R$^2$ and RELRMSE. The best performances reveal mean R$^2$ values of 0.830 and 0.783 for GPR and NCA, respectively; whereas the performance of the nonlinear methods tends to a slight decrease when the number of features reaches nine for the GPR method.

Evaluation of all feature selection results in terms of $ R^2 $ and RELRMSE values reveal that nine features designated as important by the GPR method are as follows: $AR$, $l_w$, $t_w$, $f_{c}^{'}$, ${{\rho }_{l}}$, ${{\rho }_{bl}}$, ${P}/{({{A}_{g}}}f_{c}^{'})$, $b_0$, and ${s}/{d}_{b}$. By decreasing the number of features, the complexity of the predictive model is reduced and a more practical use is achieved.

 Table \ref{table:3} shows feature weights of the nine most influential features indicated by the GPR method based on regression analyses to estimate NCDE values as well as their feature weights based on the NCA method. The observations can be summarized as: i) the GPR method gives different ranking of features than the NCA method (which is expected as each method consider different criteria in selecting the features), ii) wall length (${l}_{w}$) is identified as the most influence feature in both methods,  and iii) longitudinal web reinforcement (${{\rho }_{l}}$) and the ${s}/{d}_{b}$ have no influence on NCDE based on the NCA method.

\begin{table}[h!]
\fontsize{4}{4}
\caption{Feature weights of the best predictive models based on GPR and NCA methods for NCDE.}
\centering
\begin{tabular}{l l l l l l l l l l} 
 \hline \hline 
 \textbf{METHODS} & \textbf{$AR$} &  \textbf{$l_w$} &  \textbf{$t_w$} & \textbf{$f_{c}^{'}$} & \textbf{${{\rho }_{l}}$} & \textbf{${{\rho }_{bl}}$} & \textbf{${P}/{({{A}_{g}}} f_{c}^{'})$} & \textbf{$b_0$} & \textbf{${s}/{d}_{b}$}\\ [0.5ex] 
 \hline\hline
  GPR &  0.156 &  0.193 &  0.117 & 0.064 & 0.035 & 0.100 & 0.066 & 0.143 & 0.127 \\ 
  NCA &  0.176 &  0.190 &  0.160 & 0.126 & - & 0.112 & 0.137 & 0.097 & - \\ [0.5ex] 
 \hline
\end{tabular}
\label{table:3}
\end{table}

The features suggested by the GPR method are examined with regards to their physical meaning as follows: 

    i) Geometric properties: Experimental evidence show that shear walls have different dominant behaviors and can be classified based on their aspect ratio \cite{american2017asce}. Squat shear walls with low aspect ratio are expected to reach their shear strength and are typically damaged with diagonal tension failure or web crushing. Such walls show limited ductility which can be attributed to inelastic deformations leading to web crushing failures \cite{oesterle1981reinforcement}, thus, are expected to dissipate less energy. Similarly, slender shear walls with high aspect ratios typically show flexural yielding before reaching their shear strength and fail in flexure with damages such as concrete spalling and crushing and/or rebar buckling at the boundary elements. Such walls are expected to reach higher deformation levels where bigger loops are produced, thus, would dissipate higher energy. Wall thickness, on the other hand, has been shown to influence the energy dissipation in the literature \cite{hube2014seismic}.

    ii) Reinforcement: Longitudinal boundary reinforcement ratio (${{\rho }_{bl}}$) also has a positive impact on deformation capacity \cite{deger2019empiricalduct} as yielding of the longitudinal boundary reinforcement suppresses premature shear failure and increases ductility. Although literature review suggests that energy dissipation is insensitive to longitudinal web reinforcement (${{\rho }_{l}}$) \cite{sittipunt1995influence,hidalgo2002seismic}, the machine learning model has captured a correlation. The ${s}/{d}_{b}$ is related to the the boundary confinement ratio, and shown to have a significant influence on deformation capacity, thus, energy dissipation capacity in the literature \cite{tasnimi2000strength}.
  
    iii) High axial load ratio (${P}/{({{A}_{g}}}f_{c}^{'})$) and concrete compressive strength ($f_{c}^{'}$) has been shown to have a significant impact on the energy dissipation capacity in the literature, as summarized in Section 1. 

Therefore, it can be concluded that the wall design properties that are identified to be effective on energy dissipation capacity are physically meaningful and consistent with the experimental evidence discussed in the literature review.

\subsection{Predictive Model}

Various regression methods are utilized to provide predictive models based on the nine most influential features discussed earlier. Based on a thousand trials, the mean R$^2$ values are obtained as 0.335, 0.338, 0.771, and 0.830 for LR, LASSO, NCA, and GPR, respectively (Table \ref{table:4}). The standard deviation of R$^2$ for GPR and NCA methods are 0.083 and 0.080, respectively, indicating that the models created with these methods were robust. Table \ref{table:4} also shows that nonlinear methods yield the best outcomes in terms of the MAE, RMSE, and mean of R$^2$ values whereas linear methods are relatively poor. 

\begin{table}[h!]
\fontsize{5}{5}
\caption{Machine learning methods model results in the testing set for normalized cumulative dissipated energy (1000 random repeats).}
\centering
\begin{tabular}{l l l l l l} 
 \hline \hline 
 \textbf{METHODS} &  \begin{tabular}[c]{@{}l@{}}\textbf{Min. MAE}\\ \end{tabular} &  \begin{tabular}[c]{@{}l@{}}\textbf{Min. RMSE}\\ \end{tabular} & \begin{tabular}[c]{@{}l@{}}\textbf{The Best R${}^{2}$}\\\end{tabular} & \begin{tabular}[c]{@{}l@{}}\textbf{Mean of R${}^{2}$}\\ \end{tabular}  & \begin{tabular}[c]{@{}l@{}}\textbf{Standard}\\ \textbf{deviation of R${}^{2}$}\end{tabular}  \\
 \hline\hline
  LR & 106.727  & 154.298  &  0.768 &  0.339 & 0.176 \\ 
  LASSO &  101.522 &  144.830 &  0.740 &  0.338 & 0.155 \\
  NCA &  52.859 &  75.771 &  0.940 &  0.771 & 0.081 \\
  GPR &  51.434 &  80.669 & 0.963 &  0.830 & 0.083 \\[0.5ex] 
 \hline
\end{tabular}
\label{table:4}
\end{table}

\begin{figure}[!h]
\centering
\includegraphics[width=0.45\columnwidth]{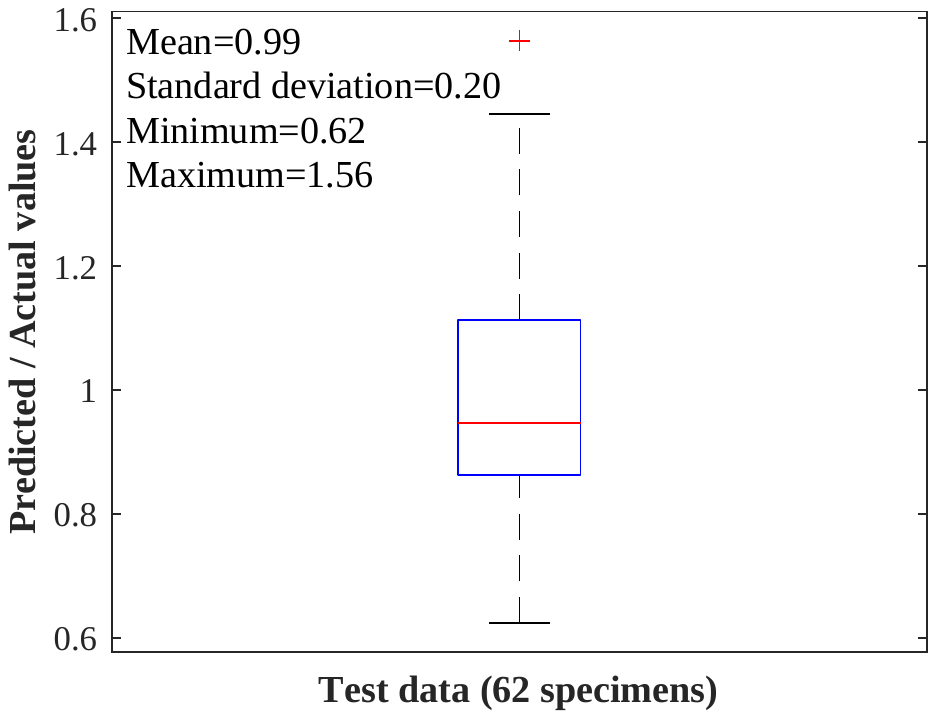}
\caption{Box chart of the ratio of predicted value to actual value for test data (results are for the repeat with R$^{2} = 0.93$).}
\label{fig:9}
\end{figure}

The GPR-based predictive model outperformed the other models with its high accuracy, therefore, is developed to estimate NCDE values. The prediction accuracy of the GPR-based model is assessed based on the ratio of predicted to actual NCDE values calculated for the test data set, revealing a mean prediction accuracy of 1.08 for a thousand random repeats; whereas the best, thus proposed, predictive model revealed mean and standard deviation of the predicted to actual NCDE values as 0.99 and 0.20, respectively. Fig. \ref{fig:9} presents a summary of statistics for the ratio of predicted to actual NCDE values, indicating that the median is 0.95 as shown by the red line. Whereas the bottom edge indicates the 25th percentile, the top edge indicates the 75th percentile. Besides, the outliers are shown using the ‘+’ symbol, and dashed lines extend to the maximum and minimum data points but except outliers.

\begin{figure}[!h]
\centering
\includegraphics[width=0.5\columnwidth]{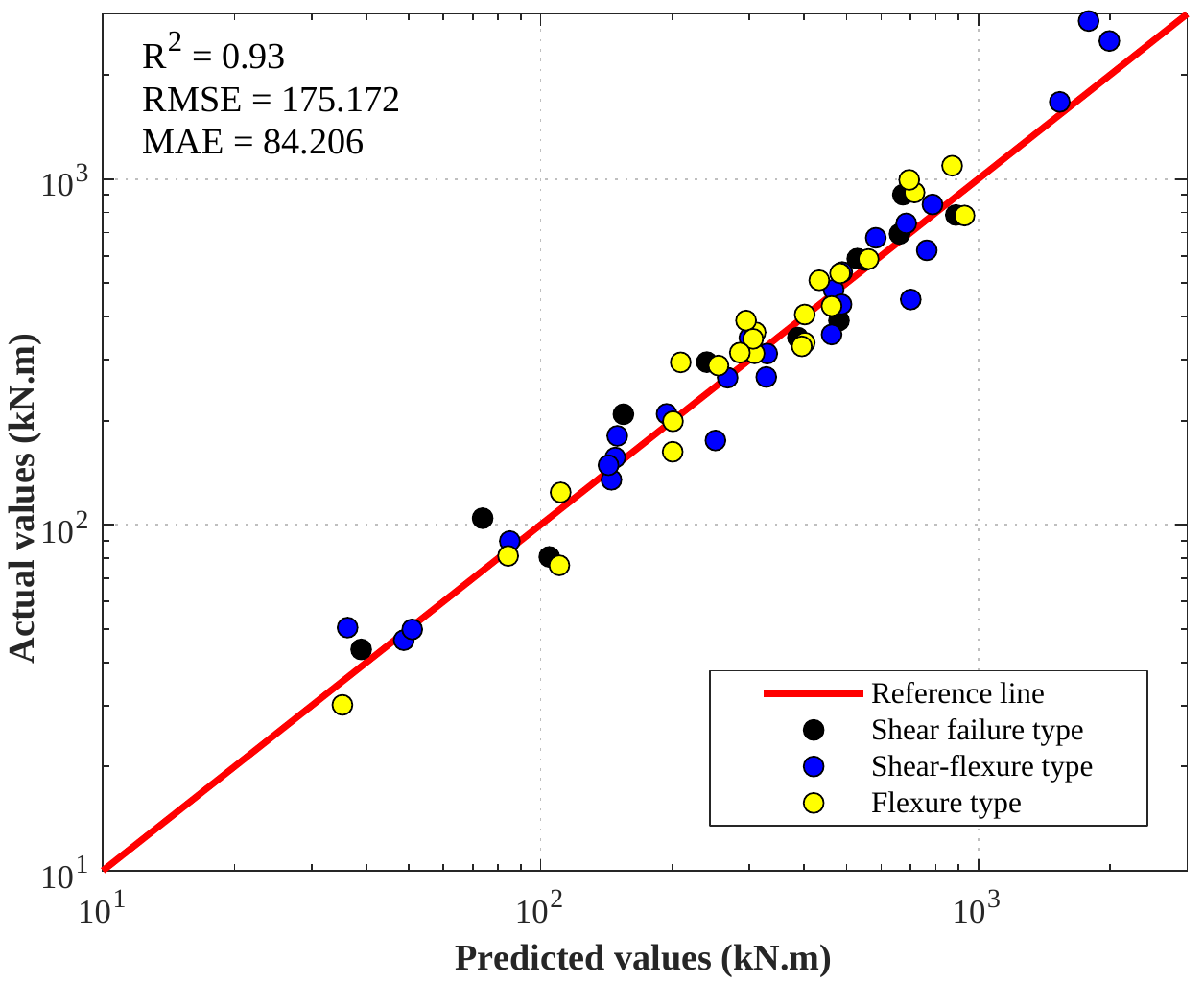}
\caption{Results of the proposed GPR-based model in the testing set for 62 shear walls.}
\label{fig:10}
\end{figure}

Fig. \ref{fig:10} compares the NCDE values predicted by the proposed model to the experimental results. Despite the large range of NCDE values, the proposed GPR-based model provides good predictions for NCDE values. To assess performance of the proposed model for shear walls with different failure modes, the test data is classified into three failure types, namely: shear, shear-flexure, and flexure failure based on the reported failure modes. The R$ ^ 2 $ values are obtained as 0.912, 0.940, and 0.899 for walls that failed in shear (12 specimens), shear-flexure (26 specimens), and flexure (24 specimens), respectively. The mean ratio of the predicted to actual NCDE values are 0.96, 1.01, and 0.98 for walls that failed in shear, shear-flexure, and flexure, respectively. It is noted that the proposed model is more reliable for rectangular shear walls compared to non-rectangular (e.g. barbell-shaped, T-shaped) walls due to the scarcity of the latter data.

\section{Summary and Conclusions}
\label{S:6}
In this study, a database consisting of 312 specimens tested worldwide is constructed to correlate the wall energy dissipation with key design parameters based on machine learning methods. The database consists of 312 reinforced concrete shear walls with rectangular (219), barbell-shaped (51), and flanged (42) sections. The cumulative energy dissipated by reinforced concrete shear wall specimens is calculated based on the experimental results and is normalized by the drift ratio at each cycle to compensate the varying number of cycles in different loading protocols. The normalized cumulative dissipated energy (NCDE) is designated as the output, whereas eighteen wall design parameters are considered as inputs (features) for the machine-learning experiments.

\begin{itemize}
    \item \underline{Feature selection:} Regression based feature selection methods reveal that nine wall design parameters (features) have relatively stronger influence on the NCDE of the shear walls, namely: wall geometric properties (i.e. length, thickness, boundary width ($b_0$), aspect ratio), concrete compressive strength, axial load ratio, longitudinal boundary and web reinforcement ratio, and transverse boundary reinforcement spacing. Based on the feature weights, geometric properties (i.e. wall length, thickness, aspect ratio, boundary width) are identified as the most effective wall properties on the energy dissipation capacity. The feature selection process not only allow to observe the wall design properties to which the NCDE is sensitive, but also help to reduce the complexity of the predictive model, thus, provide a more practical model to estimate of the energy dissipation capacity without compromising the high accuracy. The features pointed out to be influential are shown to be physically meaningful and consistent with the experimental results.
 \item \underline{Machine learning methods:}  Linear Regression (LR), Lasso Regression (LASSO), Neighborhood Component Analysis (NCA), and Gaussian Process Regression (GPR) are used to develop a model to predict the NCDE. A nonlinear relationship is found to be more suitable to represent a relation between the inputs and the output; thus the GPR method is recommended to estimate the normalized cumulative dissipated energy. 
 \item \underline{Reliability: }The reliability of the predictive model is evaluated based on the following four statistical metrics: coefficient of determination, mean absolute error, root mean square error, and mean prediction accuracy (the ratio of predicted to actual NCDE values). The mean R${}^{2}$ of the GPR method for model trials created with a thousand different train-test data sets is 0.830, whereas the R${}^{2}$ of the best trial is obtained 0.963. The ratio of predicted to actual NCDE values is very close to 1.0 indicating that high accuracy and robustness is achieved.
 \item \underline{Design comparison:} The outcomes of this study enables comparing the energy dissipation capacity of shear walls utilizing a ML-based predictive model. Due to their relatively high in-plane lateral stiffness, reinforced concrete shear walls are effective in reducing structural lateral displacement, while resisting high seismic forces and dissipating substantial energy. Energy serves as an indicator of structural damage associated with plastic deformations under cyclic loading and is a promising index for  future earthquake engineering when the energy-based approach becomes widely employed. The findings of this research are expected to resolve some of the unknown aspects of the energy-based approach. From the design point of view, the energy dissipation capacity of shear walls will be available once the input energy (demand) is known. The results of this study are also valuable in that different design (including detailing and configuration) alternatives of shear walls can be compared and evaluated based on their energy dissipation capacity. From the damage point of view, the findings of this study are indicators of structural damage in existing buildings and provide information regarding energy-based damage assessment.

\end{itemize}

\section*{Acknowledgement}
The project has been supported by funds from the Scientific and Technological Research Council of Turkey (TUBITAK) under Project No: 218M535. Opinions, findings, and conclusions in this paper are those of the authors and do not necessarily represent those of the funding agency.

\bibliographystyle{unsrt}
\bibliography{main312ZD}

\begin{thebibliography}{10}

\bibitem{housner1958efffct}
George~W. Housner and Hannu Outinen.
\newblock The efffct of torsional oscillations on earthquake stresses.
\newblock {\em Bulletin of the Seismological Society of America},
  48(3):221--229, 1958.

\bibitem{mckevitt1980hysteretic}
W.E. McKevitt, D.L. Anderson, and S.~Cherry.
\newblock Hysteretic energy spectra in seismic design.
\newblock In {\em Proceedings of the 2nd World Conference on Earthquake
  Engineering}, volume~7, pages 487--494, 1980.

\bibitem{akiyama1988earthquake}
Hiroshi Akiyama.
\newblock Earthquake resistant design based on the energy concept.
\newblock In {\em Proceedings of 9th WCEE}, pages 905--910, 1988.

\bibitem{benavent2007energy}
Amadeo Benavent-Climent.
\newblock An energy-based damage model for seismic response of steel
  structures.
\newblock {\em Earthquake engineering \& structural dynamics},
  36(8):1049--1064, 2007.

\bibitem{zahrah1984earthquake}
Tony~F. Zahrah and William~J. Hall.
\newblock Earthquake energy absorption in sdof structures.
\newblock {\em Journal of structural Engineering}, 110(8):1757--1772, 1984.

\bibitem{khashaee2003distribution}
Payam Khashaee, John~L. Gross, Payam Khashaee, Hai~Sang Lew, Bijan Mohraz, and
  Fahim Sadek.
\newblock {\em Distribution of earthquake input energy in structures}.
\newblock Diane Publishing Company, 2003.

\bibitem{benavent2010design}
A.~Benavent-Climent, Francisco L{\'o}pez-Almansa, and Diego~Andr{\'e}s
  Bravo-Gonz{\'a}lez.
\newblock Design energy input spectra for moderate-to-high seismicity regions
  based on colombian earthquakes.
\newblock {\em Soil dynamics and earthquake engineering}, 30(11):1129--1148,
  2010.

\bibitem{okur2012adaptation}
A.~Okur and M.A. Erberik.
\newblock Adaptation of energy principles in seismic design of turkish rc frame
  structures. part i: Input energy spectrum.
\newblock In {\em Proceedings of the 15th World Conference on Earthquake
  Engineering, September}, pages 24--28, 2012.

\bibitem{goel1997seismic}
Rakesh~K. Goel.
\newblock Seismic response of asymmetric systems: energy-based approach.
\newblock {\em Journal of Structural Engineering}, 123(11):1444--1453, 1997.

\bibitem{akbas2001energy}
B.~Akbas, J.~Shen, and H\_ Hao.
\newblock Energy appproach in peformance-based seismic design of steel moment
  resisting frames for basic safety objective.
\newblock {\em The structural design of tall buildings}, 10(3):193--217, 2001.

\bibitem{iancovici2008energy}
M.~Iancovici, C.~Gavrilescu, and C.~Stamatiade.
\newblock Energy distribution in high-rise buildings subjected to vrancea long
  period ground motions.
\newblock In {\em Proceedings of the 14th World Conference on Earthquake
  Engineering, China, Beijing}, pages 12--17, 2008.

\bibitem{zhang2013effects}
Sherong Zhang and Gaohui Wang.
\newblock Effects of near-fault and far-fault ground motions on nonlinear
  dynamic response and seismic damage of concrete gravity dams.
\newblock {\em Soil Dynamics and Earthquake Engineering}, 53:217--229, 2013.

\bibitem{bertero1992issues}
VITELMO~V. Bertero and CHIA-MING Uang.
\newblock Issues and future directions in the use of an energy approach for
  seismic resistant design of structures.
\newblock In {\em Fajfar P., Krawinkler H., Nonlinear seismic analysis and
  design of reinforced concrete buildings}, pages 3--22. Elsevier Applied
  Science London and New York, 1992.

\bibitem{nagae2011design}
T.~Nagae.
\newblock {\em Design and Instrumentation of the 2010 E-Defense Four-story
  Reinforced Concrete and Post-tensioned Concrete Buildings}.
\newblock PEER report. Pacific Earthquake Engineering Research Center, 2011.

\bibitem{sengupta2014hysteresis}
Piyali Sengupta and Bing Li.
\newblock Hysteresis behavior of reinforced concrete walls.
\newblock {\em Journal of Structural Engineering}, 140(7):04014030, 2014.

\bibitem{belmouden2007analytical}
Youssef Belmouden and Pierino Lestuzzi.
\newblock Analytical model for predicting nonlinear reversed cyclic behaviour
  of reinforced concrete structural walls.
\newblock {\em Engineering Structures}, 29(7):1263--1276, 2007.

\bibitem{zhang2000seismic}
Yunfeng Zhang and Zhihao Wang.
\newblock Seismic behavior of reinforced concrete shear walls subjected to high
  axial loading.
\newblock {\em Structural Journal}, 97(5):739--750, 2000.

\bibitem{yun2004behaviour}
H.-D. Yun, C.-S. Choi, and L.-H. Lee.
\newblock Behaviour of high-strength concrete flexural walls.
\newblock {\em Proceedings of the Institution of Civil Engineers-Structures and
  Buildings}, 157(2):137--148, 2004.

\bibitem{greifenhagen2005static}
Christian Greifenhagen and Pierino Lestuzzi.
\newblock Static cyclic tests on lightly reinforced concrete shear walls.
\newblock {\em Engineering Structures}, 27(11):1703--1712, 2005.

\bibitem{su2007seismic}
R.K.L. Su and S.M. Wong.
\newblock Seismic behaviour of slender reinforced concrete shear walls under
  high axial load ratio.
\newblock {\em Engineering Structures}, 29(8):1957--1965, 2007.

\bibitem{dazio2009quasi}
Alessandro Dazio, Katrin Beyer, and Hugo Bachmann.
\newblock Quasi-static cyclic tests and plastic hinge analysis of rc structural
  walls.
\newblock {\em Engineering Structures}, 31(7):1556--1571, 2009.

\bibitem{li2015experimental}
Bing Li, Zuanfeng Pan, and Weizheng Xiang.
\newblock Experimental evaluation of seismic performance of squat rc structural
  walls with limited ductility reinforcing details.
\newblock {\em Journal of Earthquake Engineering}, 19(2):313--331, 2015.

\bibitem{cardenas1973design}
Alex~E. Cardenas, John~M. Hanson, W.~Gene Corley, and Eivind Hognestad.
\newblock Design provisions for shear walls.
\newblock {\em ACI Journal}, 70(3):221--230, 1973.

\bibitem{chiou2004behavior}
Y.J. Chiou, Y.L. Mo, F.P. Hsiao, Y.W. Liou, and M.S. Sheu.
\newblock Behavior of high seismic performance walls.
\newblock In {\em 13th World Conference on Earthquake Engineering}, page 3180,
  Vancouver,B.C., Canada, 2004.

\bibitem{shaingchin2007influence}
Somboon Shaingchin, Panitan Lukkunaprasit, and Sharon~L. Wood.
\newblock Influence of diagonal web reinforcement on cyclic behavior of
  structural walls.
\newblock {\em Engineering structures}, 29(4):498--510, 2007.

\bibitem{deng2008experimental}
Mingke Deng, Xingwen Liang, and Kejia Yang.
\newblock Experimental study on seismic behavior of high performance concrete
  shear wall with new strategy of transverse confining stirrups.
\newblock In {\em Proceeding of the 14th World Conference on Earthquake
  Engineering, Xi’an University of Architecture \& Technology, China}, pages
  1--8, 2008.

\bibitem{hube2014seismic}
M.A. Hube, A~Marihu{\'e}n, JC~De~la Llera, and Bozidar Stojadinovic.
\newblock Seismic behavior of slender reinforced concrete walls.
\newblock {\em Engineering Structures}, 80:377--388, 2014.

\bibitem{liu2004effect}
Hui Liu.
\newblock {\em Effect of concrete strength on the response of ductile shear
  walls}.
\newblock PhD thesis, McGill University, Montréal, Canada, 2004.

\bibitem{yan2008seismic}
S.~Yan, L.F. Zhang, and Y.G. Zhang.
\newblock Seismic performances of high-strength concrete shear walls reinforced
  with high-strength rebars.
\newblock In {\em 11th Biennial ASCE Aerospace Division International
  Conference on Engineering, Science, Construction, and Operations in
  Challenging Environments}, pages 1--8. Earth \& Space, March 3-5, 2008.

\bibitem{layssi2012seismic}
Hamed Layssi, William~D Cook, and Denis Mitchell.
\newblock Seismic response and cfrp retrofit of poorly detailed shear walls.
\newblock {\em Journal of Composites for Construction}, 16(3):332--339, 2012.

\bibitem{sittipunt1995influence}
Chadchart Sittipunt and Sharon~L. Wood.
\newblock Influence of web reinforcement on the cyclic response of structural
  walls.
\newblock {\em Structural Journal}, 92(6):745--756, 1995.

\bibitem{hidalgo2002seismic}
Pedro~A. Hidalgo, Christian~A. Ledezma, and Rodrigo~M. Jordan.
\newblock Seismic behavior of squat reinforced concrete shear walls.
\newblock {\em Earthquake Spectra}, 18(2):287--308, 2002.

\bibitem{clough1966effect}
Ray~W. Clough.
\newblock Effect of stiffness degradation on earthquake ductility requirements.
\newblock In {\em Proceedings of Japan earthquake engineering symposium},
  page~75, Univ. of California, Berkeley, 1966.

\bibitem{takeda1970reinforced}
Toshikazu Takeda, Mete~Avni Sozen, and N.~Norby Nielsen.
\newblock Reinforced concrete response to simulated earthquakes.
\newblock {\em Journal of the Structural Division}, 96(12):2557--2573, 1970.

\bibitem{saiidi1979simple}
Mehdi Saiidi and Mete~Avni Sozen.
\newblock Simple and complex models for nonlinear seismic response of
  reinforced concrete structures.
\newblock Technical report, University of Illinois Engineering Experiment
  Station. College of~…, 1979.

\bibitem{jeon2014statistical}
Jong-Su Jeon, Abdollah Shafieezadeh, and Reginald DesRoches.
\newblock Statistical models for shear strength of rc beam-column joints using
  machine-learning techniques.
\newblock {\em Earthquake engineering \& structural dynamics},
  43(14):2075--2095, 2014.

\bibitem{zhang2018machine}
Yu~Zhang, Henry~V. Burton, Han Sun, and Mehrdad Shokrabadi.
\newblock A machine learning framework for assessing post-earthquake structural
  safety.
\newblock {\em Structural safety}, 72:1--16, 2018.

\bibitem{davoudi2018structural}
Rouzbeh Davoudi, Gregory~R. Miller, and J.~Nathan Kutz.
\newblock Structural load estimation using machine vision and surface crack
  patterns for shear-critical rc beams and slabs.
\newblock {\em Journal of Computing in Civil Engineering}, 32(4):04018024,
  2018.

\bibitem{hall2009weka}
Mark Hall, Eibe Frank, Geoffrey Holmes, Bernhard Pfahringer, Peter Reutemann,
  and Ian~H. Witten.
\newblock The weka data mining software: an update.
\newblock {\em ACM SIGKDD explorations newsletter}, 11(1):10--18, 2009.

\bibitem{Lee2020}
Sunghoon Lee.
\newblock Using weka in matlab, (2020).
\newblock
  https://www.mathworks.com/matlabcentral/fileexchange/50120-using-weka-in-matlab/.

\bibitem{mangalathu2018classification}
Sujith Mangalathu and Jong-Su Jeon.
\newblock Classification of failure mode and prediction of shear strength for
  reinforced concrete beam-column joints using machine learning techniques.
\newblock {\em Engineering Structures}, 160:85--94, 2018.

\bibitem{luo2018machine}
Huan Luo and Stephanie~German Paal.
\newblock Machine learning--based backbone curve model of reinforced concrete
  columns subjected to cyclic loading reversals.
\newblock {\em Journal of Computing in Civil Engineering}, 32(5):04018042,
  2018.

\bibitem{huang2019classification}
Honglan Huang and Henry~V. Burton.
\newblock Classification of in-plane failure modes for reinforced concrete
  frames with infills using machine learning.
\newblock {\em Journal of Building Engineering}, 25:100767, 2019.

\bibitem{siam2019machine}
Ahmad Siam, Mohamed Ezzeldin, and Wael El-Dakhakhni.
\newblock Machine learning algorithms for structural performance
  classifications and predictions: Application to reinforced masonry shear
  walls.
\newblock In {\em Structures}, volume~22, pages 252--265. Elsevier, 2019.

\bibitem{mangalathu2020data}
Sujith Mangalathu, Hansol Jang, Seong-Hoon Hwang, and Jong-Su Jeon.
\newblock Data-driven machine-learning-based seismic failure mode
  identification of reinforced concrete shear walls.
\newblock {\em Engineering Structures}, 208:110331, 2020.

\bibitem{salonikios2000cyclic}
Thomas~N. Salonikios, Andreas~J. Kappos, Ioannis~A. Tegos, and Georgios~G.
  Penelis.
\newblock Cyclic load behavior of low-slenderness reinforced concrete walls:
  failure modes, strength and deformation analysis, and design implications.
\newblock {\em ACI Structural Journal}, 97(1):132--141, 2000.

\bibitem{mohamed2014experimental}
Nayera Mohamed, Ahmed~Sabry Farghaly, Brahim Benmokrane, and Kenneth~W. Neale.
\newblock Experimental investigation of concrete shear walls reinforced with
  glass fiber--reinforced bars under lateral cyclic loading.
\newblock {\em Journal of Composites for Construction}, 18(3):A4014001, 2014.

\bibitem{kuang2008seismic}
Jun~Shang Kuang and Y.B. Ho.
\newblock Seismic behavior and ductility of squat reinforced concrete shear
  walls with nonseismic detailing.
\newblock {\em ACI Structural Journal}, 105(2):225, 2008.

\bibitem{perus2014series}
Iztok Perus, Dionysis Biskinis, Peter Fajfar, Michael~N. FARDIS, Sofia
  Grammatikou, Helmut Krawinkler, and D.G. Lignos.
\newblock The series database of rc elements.
\newblock In {\em 2nd European Conference on Earthquake Engineering and
  Seismology, Istanbul}, pages 25--29, 2014.

\bibitem{deger2021empiricalshear}
Zeynep~Tuna Deger and Cagri Basdogan.
\newblock Empirical equations for shear strength of conventional reinforced
  concrete shear walls.
\newblock {\em ACI Structural Journal}, 118(2):61--71, 2021.

\bibitem{usta2017shear}
Merve Usta, Abdallah Alhmood, Julian Carrillo, Antoni Cladera, Lucas Laughery,
  Santiago Pujol, Aishwarya Puranam, Jeffrey Rautenberg, Halil Sezen, Lesley~H.
  Sneed, et~al.
\newblock {\em Shear Strength of Structural Walls Subjected to Load Cycles}.
\newblock PhD thesis, Purdue University, West Lafayette, IN, 2017.

\bibitem{tubitak1001report}
Zeynep~Tuna Deger, G.~Taskin~Kaya, F.~Sutcu, B.~Topaloglu, and S.~Tahaei.
\newblock Betonarme perdelerde enerji sönümleme davranisinin meta-modelleme
  yöntemleriyle İncelenmesi (in turkish).
\newblock Technical report, The Scientific and Technological Research Council
  of Turkey, Ankara,Turkey, 2021.

\bibitem{ali1991rc}
Aejaz Ali and James~K. Wight.
\newblock Rc structural walls with staggered door openings.
\newblock {\em Journal of Structural Engineering}, 117(5):1514--1531, 1991.

\bibitem{lee2010computerized}
Ying Lee and Chien-Hung Wei.
\newblock A computerized feature selection method using genetic algorithms to
  forecast freeway accident duration times.
\newblock {\em Computer-Aided Civil and Infrastructure Engineering},
  25(2):132--148, 2010.

\bibitem{shang2018imputation}
Qiang Shang, Zhaosheng Yang, Song Gao, and Derong Tan.
\newblock An imputation method for missing traffic data based on fcm optimized
  by pso-svr.
\newblock {\em Journal of Advanced Transportation}, 2018, 2018.

\bibitem{Mathwork2020}
MathWorks.
\newblock Statistics and machine learning toolbox: User's guide (r2020b), 2020.
\newblock https://se.mathworks.com/help/stats/index.html/.

\bibitem{friedman2001elements}
Jerome Friedman, Trevor Hastie, and Robert Tibshirani.
\newblock {\em The elements of statistical learning}.
\newblock Springer series in statistics New York, 2001.

\bibitem{goldberger2004neighbourhood}
Jacob Goldberger, Geoffrey~E Hinton, Sam Roweis, and Russ~R Salakhutdinov.
\newblock Neighbourhood components analysis.
\newblock {\em Advances in neural information processing systems}, 17:513--520,
  2004.

\bibitem{yang2012neighborhood}
Wei Yang, Kuanquan Wang, and Wangmeng Zuo.
\newblock Neighborhood component feature selection for high-dimensional data.
\newblock {\em JCP}, 7(1):161--168, 2012.

\bibitem{williams2006gaussian}
Christopher~K.I. Williams and Carl~Edward Rasmussen.
\newblock {\em Gaussian processes for machine learning}.
\newblock MIT press Cambridge, MA, 2006.

\bibitem{kang2020displacement}
Fei Kang and Junjie Li.
\newblock Displacement model for concrete dam safety monitoring via gaussian
  process regression considering extreme air temperature.
\newblock {\em Journal of Structural Engineering}, 146(1):05019001, 2020.

\bibitem{jiang2008can}
Yue Jiang, Bojan Cukic, and Tim Menzies.
\newblock Can data transformation help in the detection of fault-prone modules?
\newblock In {\em Proceedings of the 2008 workshop on Defects in large software
  systems}, pages 16--20, 2008.

\bibitem{box1964analysis}
George~E.P. Box and David~R. Cox.
\newblock An analysis of transformations.
\newblock {\em Journal of the Royal Statistical Society: Series B
  (Methodological)}, 26(2):211--243, 1964.

\bibitem{filliben1975probability}
James~J. Filliben.
\newblock The probability plot correlation coefficient test for normality.
\newblock {\em Technometrics}, 17(1):111--117, 1975.

\bibitem{american2017asce}
ASCE.
\newblock {\em ASCE Standard, ASCE/SEI, 41-17, Seismic Evaluation and Retrofit
  of Existing Buildings}.
\newblock ASCE.: Standard. American Society of Civil Engineers, 2017.

\bibitem{oesterle1981reinforcement}
RG~Oesterle, AE~Fiorato, and William~Gene Corley.
\newblock Reinforcement details for earthquake-resistant structural walls.
\newblock {\em Research and Development Bulletin}, 1981.

\bibitem{deger2019empiricalduct}
Zeynep~Tuna Deger and Cagri Basdogan.
\newblock Empirical expressions for deformation capacity of reinforced concrete
  structural walls.
\newblock {\em ACI Structural Journal}, 116(6), 2019.

\bibitem{tasnimi2000strength}
A.A. Tasnimi.
\newblock Strength and deformation of mid-rise shear walls under load reversal.
\newblock {\em Engineering Structures}, 22(4):311--322, 2000.

\end{thebibliography}

\end{document}